\documentclass{article}
\usepackage{a4wide,amsmath,amssymb,graphicx,bbm,color,url}

\newcommand{\assign}{:=}
\newcommand{\emdash}{---}
\newcommand{\tmem}[1]{{\em #1\/}}
\newcommand{\tmmathbf}[1]{\ensuremath{\boldsymbol{#1}}}
\newcommand{\tmop}[1]{\ensuremath{\operatorname{#1}}}
\newcommand{\tmstrong}[1]{\textbf{#1}}

\newcommand{\tmtexttt}[1]{{\ttfamily{#1}}}
\newenvironment{tmparmod}[3]{\begin{list}{}{\setlength{\topsep}{0pt}\setlength{\leftmargin}{#1}\setlength{\rightmargin}{#2}\setlength{\parindent}{#3}\setlength{\listparindent}{\parindent}\setlength{\itemindent}{\parindent}\setlength{\parsep}{\parskip}} \item[]}{\end{list}}
\definecolor{grey}{rgb}{0.75,0.75,0.75}
\definecolor{orange}{rgb}{1.0,0.5,0.25}
\definecolor{brown}{rgb}{0.5,0.25,0.0}
\definecolor{pink}{rgb}{1.0,0.5,0.5}
\newtheorem{algorithm}{Algorithm}

\begin{document}

\title{Faster exact Markovian probability functions for motif occurrences: a
DFA-only approach}
\author{
\begin{minipage}{\textwidth}
\begin{center}
Paolo Ribeca,{\hspace{0.5cm}}Emanuele
Raineri\footnote{\texttt{paolo.ribeca@gmail.com}{ {\small{(corresponding author)}}, }\texttt{emanuele.raineri@gmail.com}}\\
\vspace*{1mm}\small{Systems Biology Unit, Center for Genomic Regulation, C/Dr.Aiguader 88, E08003 Barcelona, Spain}
\end{center}
\end{minipage}
}

\maketitle

\begin{abstract}
  {\tmstrong{Background:}} The computation of the statistical properties of
  motif occurrences has an obviously relevant practical application: for
  example, patterns that are significantly over- or under-represented in the
  genome are interesting candidates for biological roles. However, the problem
  is computationally hard; as a result, virtually all the existing pipelines
  (for instance {\cite{trawler}})
  use fast but approximate scoring functions, in spite of the
  fact that they have been shown to systematically produce incorrect results
  {\cite{nuelfmci}}{\cite{nuelstats}}. A few interesting exact approaches are 
  known {\cite{nuelfmci}}{\cite{zhang}}, but they are very slow and hence not
  practical in the case of realistic sequences. {\tmstrong{Results:}} We give
  an exact solution, solely based on deterministic finite-state automata
  (DFAs), to the problem of finding not only the $p$-value, but the whole
  relevant part of the Markovian probability distribution function of a motif
  in a biological sequence. In particular, the time complexity of the
  algorithm in the most interesting regimes is far better than that of
  {\cite{nuelfmci}}, which was the fastest similar exact algorithm known
  to date; in many cases, even approximate methods are outperformed.
  {\tmstrong{Conclusions:}} DFAs are a standard tool of computer science for
  the study of patterns, but so far they have been sparingly used in the study
  of biological motifs. Previous works {\cite{nuelfmci}}{\cite{proteome}}
  do propose algorithms involving automata, but there they are used
  respectively as a first step to build a Finite Markov Chain Imbedding
  (FMCI), or to write a generating function: whereas we only rely on the
  concept of DFA to perform the calculations. This innovative approach can
  realistically be used for exact statistical studies of very long genomes and
  protein sequences, as we illustrate with some examples on the scale of the
  human genome.
\end{abstract}

\section{Introduction}

It is difficult for analysis tools to meet the challenge represented by the
ever-increasing data flux coming from high-throughput post-genomics
experiments. One sector where this problem is particularly evident is that of
sequence analysis.

As it is well known, for example, the detection of statistically relevant
nucleotide sequences that occur repeatedly in a (possibly very long) stretch
of genetic material has interesting biological aspects. Motifs appearing
significantly more or less than mere chance would dictate can hint to the
presence of relevant regulatory regions, e.g. promoters or tandem repeats.
Statistical properties may relate to structural ones as well: recently, for
example, a 117Kb periodicity in {\tmem{E.coli}} has been linked to topological
features of its chromosomes {\cite{period117}}; the exceptional statistics of
the ``{\tmem{crossover hot spot instigator}}'' close to the genome core common
to different strains of {\tmem{E.coli}} can be possibly be linked to the
protection of that part of the chromosome from recombination events
{\cite{schbathchi}}.

However, it is not easy to pinpoint what {\tmem{statistically relevant}}
exactly means in the context of sequence analysis. In particular, the problem
is computationally hard for at least two reasons:
\begin{enumerate}
  \item given a sequence of symbols taken from an alphabet $\mathcal{A}$ of
  size $a$, the number of possible sequences of length $L$ is $a^L$, that is,
  the number of possible strings grows exponentially with the length of the
  considered stretch. It should then not come as a surprise to the reader the
  fact that, at least in one approach found in the literature, the solution is
  given in terms of an NP-hard algorithm {\cite{zhang}}.
  
  \item complicated motifs can overlap in non-trivial ways, making simple
  statistical approximations unreliable and requiring the exploitation of more
  sophisticated analytic techniques. As a matter of fact, it has been proved
  that in many common situations approximate methods do systematically fail to
  predict correct statistical estimators
  {\cite{nuelfmci}}{\cite{nuelstats}}; thus, possessing a fast
  non-approximate method would seem essential to provide solid and unbiased
  foundations to characterize under- and over-represented motifs from the
  point of view of their biological role.
\end{enumerate}
In fact, a good deal of attention has been devoted in the recent past to the
problem of detecting DNA motifs which appear with anomalous frequency inside a
genome, and this problem has prompted the development of a diverse range of
tools and algorithms for the study of the statistics of pattern occurrences
{\cite{zhang}}{\cite{robin}}{\cite{schbathbook}}. However, to
be computationally feasible most of the proposed methods involve either quite
drastic approximations on the statistical model which is used to describe the
genome (see for example {\cite{trawler}}), or some additional information
({\tmem{e.g.}}, a training set) to be supplied by the user
(see for example {\cite{HMMs}}).

Although very slow in many realistic regimes and as a consequence probably
unpractical for everyday use, a few algorithms to compute exact probability
distribution functions for motif occurrences are known. We distinguish two
main methods: the first one, based on position-weighted matrices, is presented
in {\cite{zhang}}; the second one, introduced in {\cite{nuelfmci}},
takes advantage of Finite Markov Chain Imbeddings (FMCIs) using Deterministic
Finite-state Automata (DFAs) to deduce the Markovian transition matrix of the
model.

In this paper we introduce a third exact method. The general statistical setup
is similar to that of the latter method (we take as null hypothesis the fact
that our sequence is generated by a Markov model of arbitrary order), but the
proposed algorithm is very different. In particular, we show that:
\begin{enumerate}
  \item contrarily to what all the recent literature about exactly solved
  Markov models would seem to imply, FMCIs are completely inessential to
  evaluate the probability distribution functions of motifs.
  
  \item a simpler approach entirely based on DFAs is possible. The simpler
  resulting algorithm naturally lends itself to a much better optimization,
  making the exact analysis of genomes of realistic size feasible. In fact, in
  many regimes the obtained performance is even better than that of the
  {\tmem{approximate}} large-deviations and Gaussian models of
  {\cite{nuelstats}} (see Table~\ref{Timings}).
\end{enumerate}
As an application, we produce in Section~\ref{Examples} an analysis of various
motifs in the human X chromosome ($L \sim 1.5 \cdot 10^8$), and of more than 16.000
transcription-factor binding sites in {\tmem{S.cerevisiae}} ($L \sim
1.2 \cdot 10^7$). Both cases are out of reach for exact FMCI methods (see Table~\ref{Timings}).

\subsection{Background about exact Markovian methods}

Defining $L$ as the length of the sequence under analysis, $N_{\tmop{obs}}$ as
the number of observed occurrences of the examined motif $\mathfrak{m}$, $a$
as the number of symbols in the alphabet $\mathcal{A}$, and $s$ as the number
of states of a DFA which is able to recognize and count the motif
$\mathfrak{m}$ (see Section~\ref{A case study}), the exact algorithm presented
in {\cite{nuelfmci}} to compute the $p$-value of $\mathfrak{m}$ for a
Markov model of order $m$ runs in a time proportional to
\begin{equation}
  \mathcal{O} \left( a \times s \times \left( N_{\tmop{obs}} + 1 \right)
  \times L_{_{_{}}} \right) . \label{NuelsCost}
\end{equation}
It is not immediate to get a practical comprehension of the meaning of such an
expression, so we briefly analyze it here. We can distinguish two main
relevant asymptotic regimes: the {\tmem{short-pattern}} regime, and the
{\tmem{long-pattern}} regime.

In the short-pattern regime, the cost of the algorithm turns out to be
essentially {\tmem{quadratic}} in the length $L$ of the considered sequence.
In fact, in the case of $m = 0$ and uniform probability distribution of
symbols the typical recorded number of observed pattern occurrences may be
estimated as
\begin{equation}
  N_{\tmop{obs}} \sim \frac{L}{a^{\ell}}, \label{Nobs}
\end{equation}
being $\ell$ the length of the pattern; similar results hold for more
complicated Markov models. Inserting this estimate into (\ref{NuelsCost}), one
immediately realizes that in this regime the cost becomes proportional to
$L^2$.

On the other hand, for long patterns the cost is {\tmem{linear}} in $L$; in
fact {\emdash}as suggested by (\ref{Nobs}) again{\emdash} the typical
occurrence numbers in this case are $N_{\tmop{obs}} = 0$ or $N_{\tmop{obs}} =
1$, so the cost becomes essentially independent of $N_{\tmop{obs}}$ and
proportional to $L$.

Of course, being the size of realistic biological sequences very large (for
example, in the range $10^7$-$10^{10}$ for the case of a typical genome) an
algorithm which is linear, or {\emdash}even worse{\emdash} quadratic in $L$ is
essentially unpractical: both the long- and the short-pattern regimes will be
unaccessible to it.

\subsection{Results and discussion\label{Results and discussion}}

In this paper we show how a formulation of motif counting in terms of
{\tmem{systolic DFAs}} (see Section~\ref{The systolic array}) allows us to
directly deduce an algorithm with cost $\mathcal{O} \left( a \times s \times
\left( N_{\tmop{obs}} + 1 \right) \times L \right)$, that is,
equivalent in complexity to that presented in {\cite{nuelfmci}}.
Furthermore, additional formal developments described in Section~\ref{Further
developments} make it possible to write the probability distribution function
$p\!\left( x \right)$ of the occurrences of a motif $\mathfrak{m}$ as
\begin{equation}
  p\!\left(x\right) = \tmop{Tr}\!\left(\left(\mathcal{M}\!\left(x\right)_{_{}}
  \right)^L \cdot v_0 \right), \label{pdf}
\end{equation}
where, as explained in Section~\ref{Further developments}, $v_0$ is an
$\mathbbm{R}$-valued vector of length $s$; in turn, $\mathcal{M}\!\left( x
\right)$ is an $s \times s$ sparse square matrix with $a \times s$ non-zero
elements, each of its elements being a polynomial in $x$ of degree
$\leqslant N_{\tmop{obs}}^{\max} = L / s$ (see Section~\ref{Further developments}). Of course, through their construction
$\mathcal{M}$ and $v_0$ depend parametrically both on the order $m$ of the
Markov model and on the pattern $\mathfrak{m}$ being examined; however, for
the sake of notational simplicity we will not indicate this fact explicitly in
the rest of the paper, as much as we will often drop the dependence of
$\mathcal{M}$ on $x$ as well.

There are two main possible evaluation strategies for such an expression:
\begin{enumerate}
  \item we compute $p\!\left( x \right)$ as
  \[ p\!\left( x \right) =\mathcal{M} \cdot \left( \mathcal{M} \cdot \left(
     \ldots \cdot \left( \mathcal{M} \cdot v_0 \right) \ldots \right) \right)
     ; \]
  we are able to take advantage of the sparsity of the matrix $\mathcal{M}$,
  but the resulting algorithm has a final complexity of
  \[ \mathcal{O}\!\left( a \times s \times \left( N_{\tmop{obs}} + 1 \right)^2
     \times L \right), \]
  being thus slower than the FMCI method in {\cite{nuelfmci}} due
  essentially to the quadratic cost of polynomial multiplication of matrix
  elements.
  
  \item we compute $\mathcal{M}^L$ directly by logarithmic decomposition, and
  its product with the initial condition vector $v_0$ in the end. The naive
  cost of such a scheme is now
  \[ \mathcal{O}\!\left( s^3 \times \left( N_{\tmop{obs}} + 1 \right)^2 \times
     \log L \right), \]
  which is much less than (\ref{NuelsCost}) in the linear long-pattern regime.
  
  In addition, in Section~\ref{TheAlgorithm} we observe that it is possible to
  use the FFT algorithm to perform polynomial multiplication as a convolution
  of the coefficients; as a result, the whole bulk of $p\!\left( x \right)$ may
  be obtained with a complexity of
  \begin{equation}
    \mathcal{O}\!\left( s^2 \times \log L \times L \right) \label{PreCost}
  \end{equation}
  which is much faster than that of the algorithm of {\cite{nuelfmci}} in
  the quadratic short-pattern regime. In addition, defining
  $p_{\varepsilon}\!\!\left( x \right)$
  as the truncated distribution obtained from $p\!\left( x
  \right)$ by suppressing the tail regions as long as $p\!\left( x \right) <
  \varepsilon \max p$, and introducing the quantity
  \begin{equation}
    \tmop{base}_{\varepsilon}\!\!\left( p \right) \assign \tmop{length}\!\left(
    \tmop{supp} p_{\varepsilon} \right), \label{base}
  \end{equation}
  a successive refinement of the method allows to obtain for this technique an
  even better final cost of
  \begin{equation}
    \mathcal{O}\!\left( s^3 \times \tmop{base}_{\varepsilon}\!\!\left( p \right)
    \times \log L \right), \label{FinalCost}
  \end{equation}
  which is excellent in the case of the evaluation of both the linear and the
  quadratic regime, given that $\tmop{base}_{\varepsilon}\!\!\left( p \right) \ll
  L$ in all practical cases (see Table~\ref{Timings}).
\end{enumerate}
To illustrate the quality of the results just obtained we consider two
examples from real-life situations in the case of a Markov model of order $1$:
\begin{enumerate}
  \item a very long sequence (i.e. the entire human genome, $\sim 3 \cdot
  10^9$ base pairs) and a long pattern of length $16$, such that
  $N_{\tmop{obs}} = 1$. In this case, one can deduce from (\ref{FinalCost})
  that the complexity speedup obtained by our algorithm w.r.t. the FMCI-based
  one would be $\sim 2 \cdot 10^5$.
  
  \item an entire human chromosome ($\sim 1.5 \cdot 10^8$ base
  pairs) and a very short pattern, \tmtexttt{ATC} for example; this is the
  nastiest possible case for the exact algorithm of {\cite{nuelfmci}},
  since it falls in its quadratic regime. Here our algorithm outperforms the
  FMCI one by a factor of $\sim 5 \cdot 10^6$.
\end{enumerate}
In fact, Equation~(\ref{FinalCost}) tells us that the performance of our
{\tmem{exact}} technique is quite close to that of the {\tmem{approximate}}
Gaussian method described in {\cite{nuelstats}} (which is essentially
proportional to $s \times \log L$); maybe more surprisingly, in the most
interesting regime of moderate $m$, long sequences and not too short patterns
our method is in general even faster than both the large-deviations and the
Gaussian approximations, and possibly much faster (see Table~\ref{Timings}
below). In addition, when thinking about these comparisons it should not be
forgotten that {\emdash}unlike approximate methods{\emdash} our faster scheme
is still able to produce the {\tmem{whole}} bulk of the probability
distribution, from which all the interesting statistical quantities can be
straightforwardly computed. A more complete comparison of the existing Markov
methods to ours is presented in Section~\ref{Examples}, together with a
discussion of the timings and the results obtained for some test examples.

\section{System and methods}

It is very easy to write a computer program which, given a stretch of DNA or
protein of length $L$, finds all the contained sub-patterns of any length
$\ell$ together with the corresponding frequencies. As mentioned before,
however, supposing that a given motif has been found $N_{\tmop{obs}}$ times it
is not easy to define and compute what the expected occurrence probability $p\!
\left( N_{\tmop{obs}} \right)$ should be: {\tmem{i.e.}}, it is not easy to
guess whether the measured number $N_{\tmop{obs}}$ is {\tmem{a priori}} large
or small.

In fact, a satisfactory and very well-known conceptual framework to tackle
similar problems has been formulated decades ago in the context of computer
science, where the ability of recognizing (``parsing'') symbol strings in
programs is essential; it is based on deterministic finite-state automata (DFA
for shortness). Such theoretical devices are ubiquitous and fundamental in
computer science, so we will not describe their principles here; we just
mention that the interested reader may easily find many thorough introductions
to the field in standard computer-science literature {\emdash} one classic
reference for this subject being for example {\cite{dragon}}.

What makes DFAs particularly interesting from the point of view of biological
analysis is that they can naturally be linked to Markov models of sequences.
In fact, we can formulate the Markovian null hypothesis that our stretch of
DNA or protein is completely determined by its {\tmem{statistics of order
$m$}}, {\tmem{i.e.}} by the frequency of appearance of each unique sub-string
of length $m + 1$ (with $m \geqslant 0$) contained in the sequence; in this
case, if we apply to the string being examined a sliding window of length $m +
1$, we can straightforwardly reinterpret the stretch as a sequence of
consecutive transitions between groups of $m + 1$ symbols, and link the Markov
chain statistics to the probability for the DFA to make a transition from a
state to another one.

This remarkable fact has been realized only recently in the context of the
analysis of biological sequences {\cite{nuelfmci}}; however, although
leading to the relatively fast FMCI-based algorithm mentioned before, the
approach presented in {\cite{nuelfmci}} uses DFAs just as auxiliary tools
to produce the transition matrix between states of the underlying Markov
process.

In the present article we show how the statistical description of a sequence
in terms of pure DFAs is worthwhile by itself, and may lead to much faster
numerical evaluation schemes. To make the reader more at ease, throughout the
following sections we will illustrate the formal developments of our technique
by means of a worked example in the case of statistics of order 0.

\subsection{A case study\label{A case study}}

Let us consider a stretch of genome with $L = 10$ in which the motif
\tmtexttt{ATC} appears two times. To decide if this frequency means that
\tmtexttt{ATC} is for some reason overrepresented, given some Markovian
statistics we must compute the probability that such a pattern can appear
randomly two times in a genome stretch of length $10$, and compare it to the
observed probability of the event. More in general, how to calculate the
probability of \tmtexttt{ATC} to appear $n$ times?\begin{figure}[tbh]
  \begin{center}\resizebox{15cm}{!}{\includegraphics{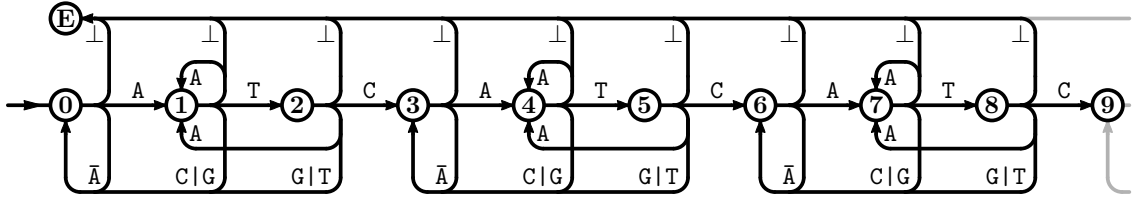}}\end{center}
  \caption{\label{ATCx3-automaton}\small An infinite DFA able to recognize and count
  all occurrences of motif \tmtexttt{ATC}. When the automaton is in states
  {\tmstrong{0}}, {\tmstrong{1}} or {\tmstrong{2}} the pattern has occurred
  exactly 0 times; when the automaton is in states {\tmstrong{3}},
  {\tmstrong{4}} or {\tmstrong{5}} the pattern has occurred 1 times; and so
  on. State {\tmstrong{E}} is the end-of-input.}
\end{figure}

Of course, one could try to give an answer in terms of direct enumeration. In
fact, this approach works quite well for very short sequences; however, the
reader can easily convince themselves that the method is both very expensive
and very difficult to generalize algorithmically as the length $L$ of the
sequence increases, and it becomes completely unpractical for the analysis of
genomes of realistic size (i.e., $L \sim 10^7$-$10^9$ bases).

To answer the question in more general terms, we begin by writing down the DFA
of Figure~\ref{ATCx3-automaton}.\begin{figure}[tbh]
  \begin{center}\resizebox{6cm}{!}{\includegraphics{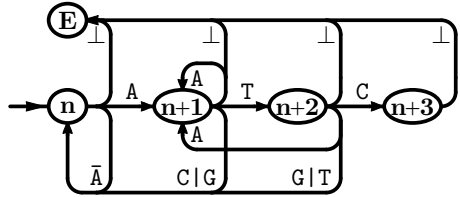}}\end{center}
  \caption{\label{ATC-automaton}\small The basic building block of Figure~\ref{ATCx3-automaton}. The complete automaton is obtained by replicating
  this block as many times as the number of observed occurrences, plus one.}
\end{figure} Such a DFA is obtained by chaining three identical basic blocks
as that represented in Figure~\ref{ATC-automaton}, each one matching one
occurrence of the original \tmtexttt{ATC} motif. Starting from state
{\tmstrong{0}}, the automaton works by reading in one character at the time
from the stretch of genome, and changing its internal state step by step
according to the input; for example, the automaton of Figure~\ref{ATCx3-automaton} when processing the string \tmtexttt{CAATCGTCATCG} will
run through the states {\tmstrong{0}}, {\tmstrong{1}}, {\tmstrong{1}},
{\tmstrong{2}}, {\tmstrong{3}}, {\tmstrong{3}}, {\tmstrong{3}},
{\tmstrong{3}}, {\tmstrong{4}}, {\tmstrong{5}}, {\tmstrong{6}},
{\tmstrong{6}}. The meaning of the states is thus as follows: when the
automaton is in state {\tmstrong{0}}, it is waiting to read the first
\tmtexttt{A}; in state {\tmstrong{1}}, it is waiting for a \tmtexttt{T} after
having read one or more \tmtexttt{A}s; in state {\tmstrong{3}}, exactly one
\tmtexttt{ATC} has been read, possibly preceded by any string different from
\tmtexttt{ATC} and followed by any number of bases different from
\tmtexttt{A}; in state {\tmstrong{6}}, exactly two \tmtexttt{ATC}s have been
read, preceded and separated by any possible string different from
\tmtexttt{ATC}, and followed by any string which is not a substring of
\tmtexttt{ATC}; and likewise for any other state (the special
character$\tmmathbf{\perp}$meaning the end of input, so that state
{\tmstrong{E}} is the final one, reached after all the input string has been
consumed by the automaton). We observe that the construction of the automaton
is far from being trivial; however, as emphasized before standard algorithms
to solve the problem are known since a long time and may easily be found in
the literature {\cite{dragon}}. Chaining more than one basic block as in
Figure~\ref{ATCx3-automaton} allows us to count the number of instances of
\tmtexttt{ATC} found in the genome: by construction, when processing any of
the genomes which contain exactly two occurrences of pattern \tmtexttt{ATC}
the automaton will always end up in states {\tmstrong{6}}, $\tmmathbf{7}$ or
$\tmmathbf{8}$.

So far, we have thus succeeded in restating our original problem: all the
genomes we are interested in ({\tmem{e.g.}} those containing two instances of
pattern \tmtexttt{ATC}) are the genomes that make the automaton of Figure~\ref{ATCx3-automaton} terminate by reaching state {\tmstrong{E}} through
states {\tmstrong{6}}, $\tmmathbf{7}$ or $\tmmathbf{8}$. How to extract direct
numerical information out of this statement? There are basically two answers
to this question.

The first answer is that by the method of generating functions one can work
out analytically the probability for the automaton to be in state
{\tmstrong{6}}, {\tmstrong{7}} or {\tmstrong{8}} (and in all other states as
well) after having read any arbitrary input; again, introductory examples to
this computational procedure may be found in {\cite{concrete}}. This method is
appealing since it allows the computation of the probability distribution
function in closed form, but it suffers from an obvious important drawback:
for a sequence of length $L$ it requires a Taylor expansion of order $L$;
given that the typical size of a relatively short bacterial genome is already
in the range of $10^6$-$10^7$ bases, one would expect this technique to be
unpractical when analyzing realistic stretches of genome. Although this is
indeed the case, it is worthwhile to note that the method has nonetheless been
exploited successfully in the context of biological research for the case of
shorter genomes {\cite{proteome}}; this result may serve as a useful
reference.

\subsection{The systolic DFA}\label{The systolic array}

We observe that a second approach is possible, which is simpler, easier to
implement and more amenable to direct interpretation and fast numerical
evaluation. Quite surprisingly, to the best of the authors' knowledge this
powerful method does not seem to have been proposed before for any relevant
practical application, neither in computer science nor in
biology.\begin{figure}[tbh]
  \begin{center}\includegraphics{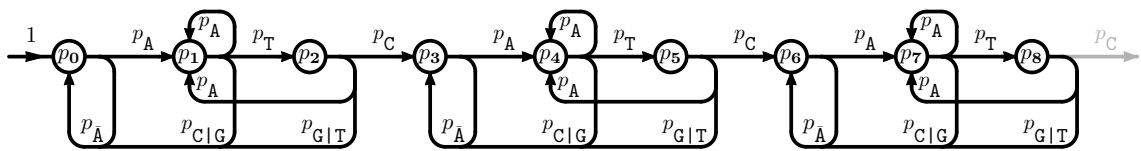}\end{center}
  \caption{\small \label{ATCx3-systolic}The automaton of Figure~\ref{ATCx3-automaton}
  reinterpreted as a systolic array. Transitions to the end-of-input state
  {\tmstrong{E}} have been omitted.}
\end{figure}

To implement it, we turn the DFA of Figure~\ref{ATCx3-automaton} into a
{\tmem{systolic array}}, which is a weighted graph where the propagation of
flow from node to node happens proportionally to the transition probabilities,
and at a constant speed of one edge per alphabet symbol read by the automaton
in the sequence; such a construction owes its name to the fact that it can be
assimilated to an hydraulic circuit where the probability $1$ enters from the
left at time $0$, and is then split and pumped at constant speed towards the
following states of the automata, reaching more and more nodes as time goes
by.

For example, let us suppose to start at time $0$ {\emdash}that is, at sequence
length $L = 0${\emdash} with all the probability in state {\tmstrong{0}}.
Since state {\tmstrong{0}} has two outgoing connections, one to state
$\tmmathbf{1}$ and one to itself, at time $1$ (or, equivalently, at length $L
= 1$) the content of state {\tmstrong{0}} will have been pumped partly into
state {\tmstrong{1}} and partly back to state {\tmstrong{0}}; in particular,
state {\tmstrong{1}} will now contain $p_{\text{{\tmstrong{1}}}}\!\left( t = 1
\right) = p_{\text{{\tmstrong{0}}}}\!\left( t = 0 \right) \cdot
p_{\text{\tmtexttt{A}}} = 1 \cdot p_{\text{\tmtexttt{A}}} =
p_{\text{\tmtexttt{A}}}$, and $p_{\text{{\tmstrong{0}}}}\!\left( t = 1 \right)$
will be the new content of state {\tmstrong{0}}, that is
$p_{\text{{\tmstrong{0}}}}\!\left( t = 0 \right) \cdot
p_{\overline{\text{\tmtexttt{A}}}} = p_{\overline{\text{\tmtexttt{A}}}}$. The
idea can easily be generalized to later times ({\tmem{i.e.}} longer sequences)
if the structure of the automaton is taken into due account.

By construction, at any given time (or sequence length $L$) the probability
distribution $p (n)$ as a function of the number of pattern occurrences $n$
may thus be obtained just by adding the contents of the states three by three:
for example, if $L = 10$ then $p(0) = p_{\tmmathbf{0}}\!\left( t = 10 \right)
+ p_{\tmmathbf{1}}\!\left( t = 10 \right) + p_{\tmmathbf{2}}\!\left( t = 10
\right)$, $p(1) = p_{\tmmathbf{3}}\!\left( 10 \right) + p_{\tmmathbf{4}}\!
\left( 10 \right) + p_{\tmmathbf{5}}\!\left( 10 \right)$, and so on (the
extension to automata with different number of states is obvious); this direct
interpretation should not be overlooked, since it is fundamental for the
developments to come. We note that the probability flows and redistributes
from node to node as a consequence of time evolution but its total amount is
not consumed in the process; in fact, the sum of the weights of the outgoing
connections for each node is by definition $1$, being it also the sum of the
transition probabilities over all possible symbols. The automaton of Figure~\ref{ATCx3-automaton} re-interpreted as a systolic array is shown in Figure~\ref{ATCx3-systolic}, and its evolution up to $t = 512$ (that is, up to $L =
512$) is listed in Table~\ref{SystolicPropagation}\begin{table}[tbh]
\begin{small}
  \begin{tabular*}{\textwidth}{c|p{3.5mm}p{3.5mm}p{3.5mm}p{3.5mm}p{2.4mm}p{3.5mm}p{3.5mm}p{3.5mm}p{6.5mm}||c|p{16mm}p{16mm}p{16mm}}
    $L$ & $p_{\tmmathbf{0}}$ & $p_{\tmmathbf{1}}$ & $p_{\tmmathbf{2}}$ &
    $p_{\tmmathbf{3}}$ & $p_{\tmmathbf{4}}$ & $p_{\tmmathbf{5}}$ &
    $p_{\tmmathbf{6}}$ & $p_{\tmmathbf{7}}$ & $p_{\tmmathbf{8}}$ &
    $L$ & $p_{\tmmathbf{0}} + p_{\tmmathbf{1}} + p_{\tmmathbf{2}}$
    & $p_{\tmmathbf{3}} + p_{\tmmathbf{4}} + p_{\tmmathbf{5}}$ &
    $p_{\tmmathbf{6}} + p_{\tmmathbf{7}} + p_{\tmmathbf{8}}$\\
    \hline
    0 & $1$ & $0$ & 0 & 0 & 0 & 0 & 0 & 0 & 0 & 16 &
    $\sim 0.797$ & $\sim 0.188$ & $\sim 0.0148$\\
    1 & $\frac{3}{4}_{_{_{}}}$ & $\frac{1}{4}$ & 0 & 0 & 0 & 0 & 0 & 0 & 0 &
    32 & $\sim 0.614$ & $\sim 0.312$ & $\sim 0.0661$\\
    2 & $\frac{11}{16}_{_{_{}}}$ & $\frac{1}{4}$ & $\frac{1}{16}$ & 0 & 0 &
    0 & 0 & 0 & 0 & 64 & $\sim 0.365$ & $\sim 0.383$ &
    $\sim 0.185$\\
    3 & $\frac{43}{64}$ & $\frac{1}{4}$ & $\frac{1}{16}$ &
    $\frac{1}{64}$ & 0 & 0 & 0 & 0 & 0 & 128 &
    $\sim 0.129$ & $\sim 0.275$ & $\sim 0.282$\\
    ... &  &  &  &  &  &  &  &  &  & 256 &
    $\sim 0.0160$ & $\sim 0.0691$ &
    $\sim 0.146$\\
    8 & $\frac{40531}{65536}$ & $\frac{3379}{16384}$ & $\frac{3841}{65536}$ &
    $\frac{2221}{32768}$ & $\frac{157}{8192}$ & $\frac{127}{32768}$ &
    $\frac{83}{65536}$ & $\frac{3}{16384}$ & $\frac{1}{65536}$ & 512 &
    $\sim 0.000249$ & $\sim 0.00215$ & $\sim 0.00922$
  \end{tabular*}
\end{small}
  \caption{\label{SystolicPropagation}\small Flow of probability propagating through
  the systolic array of Figure~\ref{ATCx3-systolic}, as the array consumes
  longer and longer stretches of genome. To compute this table we chose
  $p_{\text{\tmtexttt{A}}} = p_{\text{\tmtexttt{C}}} = p_{\text{\tmtexttt{G}}}
  = p_{\text{\tmtexttt{T}}} = 0.25$. Apart from the probability of recording
  zero patterns, which always decreases exponentially after a transient phase
  (see Section~\ref{Further developments}), the other probabilities show a
  bell-shaped behaviour w.r.t. genome length: the probability of recording
  exactly $N_{\tmop{obs}}$ patterns grows, peaks for a certain length of the
  genome stretch, and decreases afterwards. It should be noted that starting
  from $L > 8$ the probabilities add up to a number which is smaller and
  smaller than $1$, since more and more probability flux is transmitted to
  states $p_{\tmmathbf{9}}, p_{\tmmathbf{1}\tmmathbf{0}}, \ldots$ .}
\end{table}.

The interesting point about such a construction is that it is able to keep
track of the effects of {\tmem{all}} possible input sequences at the same time
by superposition, as a consequence of the fact that the probability present in
a node at time $t$ splits over all possible transitions at time $t + 1$.
Basically, we have turned our original automaton, which was a simple
recognizer, into a much more powerful device, which can now compute the
probability of being in each state after having read all possible inputs; it
should thus not come as a surprise the statement that we are now able to scan
in polynomial time sets of strings whose cardinality grows exponentially with
the length of the sequence.

We note that even a naive interpretation of this setup immediately yields a
viable computational scheme.

\begin{algorithm}
{\tmstrong{(Systolic Naive).}} Given a Markov model of order $m$
and a motif\/ $\mathfrak{m}$, compute $p\!\left( x \right)$ as follows:
\begin{enumerate}
  \item generate the systolic automaton associated to $\mathfrak{m}$ by the
  given Markov model; in particular, set a suitable initial condition
  $p_{\tmmathbf{i}}\!\left( t = 0 \right)$ for the states $(1 \leqslant i
  \leqslant s)$
  \item {\tmstrong{for}}\/ $t = 1$ {\tmstrong{to}} $L$ {\tmstrong{do}}
    \begin{tmparmod}{0pt}{0pt}{1.5em}
      evaluate the systolic propagation in the automaton:

      {\tmstrong{for}}\/ $i = 1$ {\tmstrong{to}} $s$ {\tmstrong{do}}
      \begin{tmparmod}{0pt}{0pt}{3em}
        $p_{\tmmathbf{i}}\!\left( t \right) = 0$
      \end{tmparmod}

      {\tmstrong{done}}

      {\tmstrong{foreach}} connection from state $\tmmathbf{s}_{\tmop{source}}^k$
      to state $\tmmathbf{s}_{\tmop{sink}}^k$ with weight $w_k$ $(1 \leqslant k
      \leqslant a \times s)$ {\tmstrong{do}}
      \begin{tmparmod}{0pt}{0pt}{3em}
        $p_{\tmmathbf{s}_{\tmop{sink}}^k}\!\!\left( t \right) =
        p_{\tmmathbf{s}_{\tmop{sink}}^k}\!\!\left( t \right) + w_k \cdot
        p_{\tmmathbf{s}_{\tmop{source}}^k}\!\!\left( t - 1 \right)$
      \end{tmparmod}

      {\tmstrong{done}}
    \end{tmparmod}

    {\tmstrong{done}}
  \item return the probability distribution function as
  $\displaystyle p\!\left( x \right) =
  \sum_{n = x \cdot s}^{\left( x + 1 \right) \cdot s - 1} p_{\tmmathbf{n}}\!
  \left( L \right),\;\; \forall x \in \left[ 0, N_{\tmop{obs}}^{\max} \right]$.
  $\Box$
\end{enumerate}
\end{algorithm}
This algorithm possesses at least two very desirable properties:
\begin{enumerate}
  \item apart from possible small roundoff errors in the evaluation of
  Markovian statistics, it is exact and numerically stable.
  
  \item it allows the optimized evaluation of $p\!\left( x \right)$ in the
  region $x \in \left[ 0, N_{\tmop{obs}}^{\max} \right]$; in fact, this effect
  may be obtained just by truncating the automaton after $\left(
  N_{\tmop{obs}} + 1 \right)$ basic blocks, and letting the probability flux
  which should go to higher stages simply disperse.
\end{enumerate}
Considering that each state of the automaton has $a$ possible outgoing
transitions, the computational cost of the algorithm in the latter truncated
case can readily be written as $\mathcal{O}\!\left( a \times s \times L \times
\left( N_{\tmop{obs}} + 1 \right)_{} \right)$, which by a striking coincidence
turns out to be exactly the same as that of the very different FMCI-based
exact algorithm presented in {\cite{nuelfmci}}. We also point out that
taking advantage of the cutoff technique which will be explained in Section~\ref{TheAlgorithm} we could slightly improve on this result by discarding the
tail of very small elements coming ahead of the distribution.

\subsection{Further developments\label{Further developments}}

In fact, much better results may be obtained; to proceed, however, we have to
develop a slightly more elaborate description of the problem.

\begin{figure}[tbh]
  \begin{center}\includegraphics{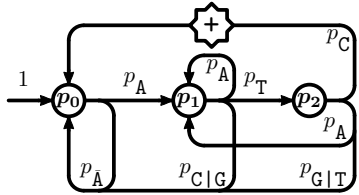}\end{center}
  \caption{\label{ATC-folded}\small The folded version of the systolic array of
  Figure~\ref{ATCx3-systolic}. The transition from state $\tmmathbf{2}$ to
  state $\tmmathbf{0}$ is a counting one, as indicated by the
  ``{\tmstrong{+}}'' symbol. Transitions to the end-of-input state have been
  omitted.}
\end{figure}

First of all, we observe that it is not really necessary to have in the
automaton as many building blocks as the largest number of pattern occurrences
$N_{\tmop{obs}}^{\max} = L / s$ which might appear in a sequence; in fact, it
is possible to get a ``folded'' version of the automaton just by adding one
single connection to its basic block. For example, the reader may easily
convince themselves that the automaton of Figure~\ref{ATC-folded} is entirely equivalent to that of Figure~\ref{ATCx3-systolic},
provided that we supply a mechanism to keep track of the fact that the
probability flux associated to a number $N$ of motif occurrences in state
$\tmmathbf{2}$ becomes associated to $N + 1$ occurrences when passing to state
$\tmmathbf{0}$ through the special connection indicated by the ``$+$'' symbol;
that is why we name such a link a {\tmem{counting transition}}. We will
describe in a moment one possible way to implement a bookkeeping mechanism
like the one just mentioned. We also note that our treatment introduces a
slight but significant difference w.r.t. the related approach presented in
{\cite{nueldfa}}: in fact, as from the standard theory of DFA construction one
finds in the latter work a final counting {\tmem{state}} instead of our
counting {\tmem{transition}}, resulting in our automaton being one state
shorter; although we will not examine further this issue in the present paper,
it is possible that our slightly more compact construction could indeed be
used to improve some of the results obtained in {\cite{nueldfa}}.

Secondly, it is worthwhile to cast the problem in matrix form. To this end, we
observe that it is possible to express the systolic propagation of the
probability flux in terms of a {\tmem{transition matrix}} $\mathcal{M}$, which
describes the modification in the content of the states of the automaton as
the computation proceeds from time $t$ to time $t + 1$ (that is, from the
length $L$ of the sequence read so far by the automaton to length $L + 1$).
This goal may be obtained quite straightforwardly just by arranging as a
matrix the transition probabilities from one state to another.

In particular, if there were not any mechanism in place to record the number
of observed patterns, $\mathcal{M}$ could be defined quite simply as an $s
\times s$ matrix of the form
\[ \mathcal{M}_{i j} \assign p\!\left( \tmmathbf{i} \rightarrow \tmmathbf{j}
   \right) . \]
For example, for the automaton in figure 4 one would write
\begin{equation}
  \mathcal{M}_0 = \left(\begin{array}{ccc}
    p_{\overline{\text{\tmtexttt{A}}}} & p_{\text{\tmtexttt{C|G}}} &
    p_{\text{\tmtexttt{C}}} + p_{\text{\tmtexttt{G|T}}}\\
    p_{\text{\tmtexttt{A}}} & p_{\text{\tmtexttt{A}}} &
    p_{\text{\tmtexttt{A}}}\\
    0 & p_{\text{\tmtexttt{T}}} & 0
  \end{array}\right), \label{Non-counting matrix}
\end{equation}
the columns representing (in order) the state which the automaton leaves on
reception of a new symbol and the rows being indexed by the state reached by
the automaton: at the intersection between any column and any row one would
find the probability of such a transition to happen. Note that by definition
all the columns would be normalized to 1.

We have only elucidated half of the matrix structure so far, though: our
finite automata has a special counting transition, which allows us to
remember, in any state, how many times the motif has been observed. A
convenient way to represent this feature is by replacing the content of each
state of the automaton with a polynomial: the coefficient of order $k$ of the
polynomial will then represent the probability that the motif has been
observed $k$ times. Accordingly to this convention, the elements of the
transition matrix need to be replaced by a polynomial as well; in particular,
one has to multiply the probability of the counting transition(s) by $x$,
since their effect is to increase by one the number of motifs met so far.

For example, assuming all the probabilities to be $p_{\text{\tmtexttt{A}}} =
p_{\text{\tmtexttt{C}}} = p_{\text{\tmtexttt{G}}} = p_{\text{\tmtexttt{T}}} =
0.25$ the complete counting transition matrix for the automaton of figure 4
will now be
\begin{equation}
  \mathcal{M}\!\left( x \right) = \left(\begin{array}{ccc}
    0.75 & 0.50 & 0.5 + 0.25 x\\
    0.25 & 0.25 & 0.25\\
    0 & 0.25 & 0
  \end{array}\right), \label{ATC-matrix}
\end{equation}
the only difference w.r.t. the non-counting matrix $\mathcal{M}_0$ of
(\ref{Non-counting matrix}) lying in the transition probability from state
{\tmstrong{2}} to state {\tmstrong{0}}, which now has been replaced by
$p_{\text{\tmtexttt{G|T}}} + p_{\text{\tmtexttt{C}}} x$.

To complete the formalism, one just needs to describe how the probability flux
is distributed among the states of the automaton at the beginning of the
computation; this can be done by simply letting the product of matrices
operate on a real-valued vector $v_0$ of length $s$ which expresses the
initial condition. The probabilities of being in any particular state can thus
be deduced by multiplying the matrix $\mathcal{M}$ by itself $L$ times and
applying the result to $v_0$; in the end, the elements of the obtained vector
must be summed over, since we are not interested any longer in how the
probability is distributed among the states of the automaton. The resulting
polynomial will then correspond to our probability distribution. This way, we
get to the final formula, Equation~(\ref{pdf}).

The reader may easily verify that a repeated application of the matrix
appearing in Equation~(\ref{ATC-matrix}) to the initial condition defined by
\[ v_0 = \left(\begin{array}{c}
     1\\
     0\\
     \vdots\\
     0
   \end{array}\right) \]
indeed reproduces Table~\ref{SystolicPropagation}. In particular, all the
construction carried out in this Section can be directly extended to the cases
of generic pattern and of Markov model of generic order by supplying the
correct automaton with its initial condition to the matrix formalism just
described (this fact is used in {\cite{nueldfa}} as well); indeed, different
choices for the model or $\mathfrak{m}$ only modify the actual contents of
$\mathcal{M}$ and $v_0$ in Equation~(\ref{pdf}), which remains correct.

This way of restating the original problem elicits very interesting
considerations. For example:
\begin{enumerate}
  \item $p_{\tmmathbf{0}}$ always decreases exponentially after a transient
  
  \item more stringent conclusions may be drawn about the unimodality of
  $p\!\left( x \right)$
  
  \item the causality requirement to the propagation of probability flux
  imposes specific constraints to the form of the transition matrix; such
  property might possibly be used to speed up the computation of $p\!\left( x
  \right)$ even more.
\end{enumerate}
In general, although in the present paper we will not pursue any formal
investigation, we emphasize that Equation~(\ref{pdf}) indeed is a perfect
place where a rigorous study of the statistical properties of Markovian
probability distribution functions can be started.

Finally, it is straightforward to obtain the naive costs of the different
possible strategies to evaluate Equation~(\ref{pdf}); we recall that such
results have already been presented in Section~\ref{Results and discussion}.

\section{Algorithm}\label{TheAlgorithm}

Everything is now in place to describe our algorithm for the fast evaluation
of probability distribution functions of Markov models.

The key observation to obtain an algorithm with superior performance w.r.t.
that presented in {\cite{nuelfmci}} is to note that {\tmem{multiplication
of polynomials may be better performed in terms of a convolution of their
coefficients}}.

\begin{algorithm}\label{FFTMultiplication}
{\tmstrong{(Polynomial multiplication by Discrete Fourier Transformation).}}
Given two polynomials $p\!\left( x \right)$ and $q\!\left( x
\right)$, compute $p q\!\left( x \right)$ as follows:
\begin{enumerate}
  \item extend $p\!\left( x \right)$ and $q\!\left( x \right)$ to two
  polynomials $p'\!\left( x \right)$ and $q'\!\left( x \right)$ of degree $d' =
  \deg\!\left( p \right) + \deg\!\left( q \right)$ by zero padding.
  
  \item define and compute $\mathfrak{F}\!\left( p q) \right.$ as\/
  $\mathfrak{F}\!\left( p q \right)_i \assign
  \mathfrak{F}\!\left( p' \right)_i \mathfrak{F}\!\left( q' \right)_i$,
  $0 \leqslant i \leqslant d'$.
  
  \item return the polynomial product as $p q\!\left( x \right)
  =\mathfrak{F}^{- 1}\!\left( \mathfrak{F}\!\left( p q \right) \right)$. $\Box$
\end{enumerate}
\end{algorithm}
Of course, the main point of interest for our application is that the naive
evaluation of $p q\!\left( x \right)$ entails $\left( \deg\!\left(p\right) + 1
\right) \times \left( \deg\!\left( q \right) + 1 \right)$ operations, while the
use of Algorithm \ref{FFTMultiplication} coupled to Fast Fourier Transform 
(FFT) would require only
$\mathcal{O}\!\left( \deg p q \times \log\!\left( \deg p q \right) \right)$
operations to perform the same computation, making the behaviour switch from
quadratic to linear. We observe that in the literature it is not easy to find
direct applications of this scheme, possibly due to its sensitivity to
numerical noise which will be analyzed in more detail below; however, the
algorithm {\tmem{is}} indeed applied in indirect form in some cases, for
example when computing products of modulo polynomials (as in the so called
{\tmem{Karatsuba multiplication}} {\cite{NR}}).

Let us note that if we consider polynomial-valued matrices the following
relation holds:
\[ \mathcal{M}\cdot\mathcal{N}=\mathfrak{F}^{- 1}\!\left( \mathfrak{F}\!\left(
   \mathcal{M}\right)\cdot\mathfrak{F}\!\left(\mathcal{N}\right)_{_{}} \right)
\]
where by definition
\[ \left( \mathfrak{F}\!\left( \mathcal{M} \right)_{_{}} \right)_{i j} \assign
   \mathfrak{F}\!\left( \mathcal{M}_{i j}\!\left( x \right)_{_{}} \right) ; \]
this means that the Fourier operator applied to polynomial coefficients
commutes with matrix multiplication.

A natural and appealing idea is then to use an FFT setup to directly compute
the power $\mathcal{M}^L$ of our polynomial-valued matrix appearing in
(\ref{pdf}). Nonetheless, two objections should be addressed.

First of all, as from Algorithm \ref{FFTMultiplication} a (technical) problem 
when using the FFT
algorithm in the context of polynomial multiplication is that care must be
taken to ensure that FFT vectors are larger than the degree of the resulting
polynomial, otherwise overlaps will occur. This issue increases the memory
requirement of the algorithm by a factor of two, but otherwise it has no
relevant effect on the evaluation strategy for $p\!\left( x \right)$.

A much more delicate issue is that the FFT algorithm is very sensitive to the
noise introduced by rounding errors; in particular, if our bell-shaped
distribution is to be represented as a (long) polynomial obtained by repeated
multiplications of shorter ones, the typical condition number of the
coefficients will be large, while the FFT algorithm can be faithful only to
the coefficients whose magnitude does not differ from that of the largest
coefficient for more than the numerical precision of the floating-point type
used; this shall have the effect of making unreliable the computation of the
smaller coefficients of the polynomial {\emdash} that is, the computation of
the tails of the distribution and hence of the $p$-value. However, it is easy
to answer this objection, at least when the proposed algorithm is used to
compute biologically relevant quantities. In detail:
\begin{enumerate}
  \item occurrences of motifs observed in practical situations only very
  rarely fall in the far tails of the distribution (and in any case, we can
  always quantify algorithmically when our computed $p$-value becomes
  unreliable).
  
  \item even if the computation of the tails becomes unreliable, the FFT
  algorithm is still perfectly able to deduce all the relevant statistical
  quantities of the distibution {\emdash}which only depend on the
  {\tmem{bulk}} of the distribution, not on the exact knowledge of its
  {\tmem{tails}}{\emdash} and hence the $z$-value, which is probably even more
  significant and informative than the $p$-value in the case of a very
  unlikely occurrence number.
\end{enumerate}
We can now formulate without concerns an effective strategy to evaluate
$\mathcal{M}^L$ in Equation~(\ref{pdf}), and thus the bulk of $p\!\left( x
\right)$, by an FFT-based technique.

\begin{algorithm}\label{SystolicFFT}{\tmstrong{(Systolic Fast via FFT).}}
Given a Markov model of order $m$ and a motif\/ $\mathfrak{m}$, compute 
$p\!\left( x \right)$ as follows:
\begin{enumerate}
  \item generate the systolic automaton associated to $\mathfrak{m}$ by the
  given model.
  
  \item deduce from the systolic automaton the transition matrix $\mathcal{M}$
  and the initial condition $v_0$.
  
  \item create the two auxiliary square matrices \tmtexttt{power} and
  \tmtexttt{result} of size $s$.
  
  \item initialize: $\text{\tmtexttt{power}} \leftarrow \mathfrak{F}\!\left(
  \mathcal{M} \right)$, $\text{\tmtexttt{result}} \leftarrow
  \mathfrak{F}\!\left( \tmop{identity}_s \right)$, $N \leftarrow L$.
  
  \item compute $\mathfrak{F}\!\left( \mathcal{M}^L \right)$ by binary
  decomposition in \tmtexttt{power} as:\\
  {\tmstrong{while}} \tmtexttt{true} {\tmstrong{do}}
  \begin{tmparmod}{0pt}{0pt}{1.5em}
  as necessary, apply a cutoff function to the elements of \tmtexttt{power}
  and\/ \tmtexttt{result}:
  
  $\text{\tmtexttt{power}} \leftarrow \tmop{cutoff} \left(
  \text{\tmtexttt{power}} \right)$, $\text{\tmtexttt{result}} \leftarrow
  \tmop{cutoff} \left( \text{\tmtexttt{result}} \right)$;
  
  {\tmstrong{if}} $N \tmop{mod} 2 = 1$ {\tmstrong{then}}
  
  \begin{tmparmod}{0pt}{0pt}{3em}
    $\text{\tmtexttt{result}} \leftarrow \text{\tmtexttt{result}}\,\cdot\,
    \text{\tmtexttt{power}};$
  \end{tmparmod}
  
  $N \leftarrow \left\lfloor N / 2 \right\rfloor$;
  
  {\tmstrong{if}} $N = 0$ {\tmstrong{then}}
  
  \begin{tmparmod}{0pt}{0pt}{3em}
    {\tmstrong{break}};
  \end{tmparmod}
  
  $\text{\tmtexttt{power}} \leftarrow \text{\tmtexttt{power}}\,\cdot\,
  \text{\tmtexttt{power}}$
  \end{tmparmod}
  {\tmstrong{done}}.
  
  \item return the probability distribution function as $\tmop{Tr}\!\left(
  \mathfrak{F}^{- 1}\!\left( \text{\tmtexttt{result}} \right) \cdot v_0
  \right)$. $\Box$
\end{enumerate}
\end{algorithm}
We will explain soon how the cutoff function appearing at point (5) should be
chosen; for the moment, let us pretend that it is the identity. In this case,
recalling that $N^{\max}_{\tmop{obs}} = L / s$ by construction of the
(systolic) automaton, the cost of the algorithm is given by
\[ \mathcal{O}\!\left( 3 \times s^2 \times \left( N_{\tmop{obs}}^{\max} + 1
   \right) \times \log\!\left( N^{\max}_{\tmop{obs}} + 1 \right) + 2 \times s^3
   \times \left( N_{\tmop{obs}}^{\max} + 1 \right) \times \log L \right)
   =\mathcal{O}\!\left( s^2 \times L \times \log L \right), \]
which is Equation~(\ref{PreCost}); the first term in the l.h.s. comes from the
Fourier transformations of $\mathcal{M}$, $\tmop{identity}_s$ and
\tmtexttt{result}, while the second one {\emdash}which is dominant if no
cutoff is applied{\emdash} is due to the $2 \times \log L$ matrix
multiplications occurring in the algorithm.

It is worthwhile to note that, although already excellent from the point of
view of performance (in particular w.r.t. the complexity of the FMCI algorithm
{\cite{nuelfmci}} in the quadratic short-pattern regime, as already
pointed out in Section~\ref{Results and discussion}) this scheme can be
further improved by an appropriate choice of the function
$\tmop{\mathit{cutoff}}$ which appears in Algorithm \ref{SystolicFFT}.

In fact, we have already pointed out before that our knowledge of $p\!\left( x
\right)$ as obtained from an FFT-based algorithm is naturally limited by the
numerical precision $\varepsilon \assign 10^{- d}$ of our floating-point type,
$d$ being the number of available decimal digits; as anticipated in Section~\ref{Results and discussion}, it is then natural to introduce a truncated
distribution function $p_{\varepsilon}\!\!\left( x \right)$, which is obtained
from $p\!\left( x \right)$ by removing its tails both on the left and the right
extrema, in the region (if any) where their value falls below the threshold
given by $\varepsilon \max p$. As argued before, $p_{\varepsilon}\!\!\left( x
\right)$ keeps all the information which are needed to compute the statistical
indicators we are interested in, and furthermore the truncation process does
not introduce numerical instabilities; however, this choice of the cutoff has
two contrasting implications on the complexity of the algorithm.

The first one is that the cutoffing step slows down the computation; in fact,
both the matrices \tmtexttt{power} and \tmtexttt{result} appearing in
Algorithm \ref{SystolicFFT} store their elements in Fourier space, and as a 
result one needs
to Fourier transform back and forth at each application of the cutoff. On the
other hand, the second consequence is that after having applied the cutoff we
need less polynomial coefficients than $N_{\tmop{obs}}^{\max} + 1 = L / s + 1$
to represent each element of $\mathcal{M}\!\left( x \right)$, since the tails
where the distribution is small have now been eliminated; this fact turns out
to be a relevant advantage in terms of computational efficiency in many cases,
winning in particular a factor which is typically very big in the large-$L$
regime, and thus justifying this choice of the cutoff.

Defining $\tmop{base}_{\varepsilon}$ as in Equation~(\ref{base}), we can then
compute the final complexity of this new algorithm as
\begin{eqnarray*}
  &  & \mathcal{O}\!\left( s^2 \times \tmop{base}_{\varepsilon}\!\!\left( p
  \right) \times \log \tmop{base}_{\varepsilon}\!\!\left( p \right) \times
  \left(3 + 4 \times \log L \right) + 2 \times s^3 \times
  \tmop{base}_{\varepsilon}\!\!\left( p \right) \times \log L \right)\\
  & = & \mathcal{O}\!\left( s^2 \times \tmop{base}_{\varepsilon}\!\!\left( p
  \right) \times \log L \times \left( 2 \times s + 4 \times \log
  \tmop{base}_{\varepsilon}\!\!\left( p \right)_{_{}} \right)_{} \right)
\end{eqnarray*}
from which Equation~(\ref{FinalCost}) immediately follows; depending on $s$
the dominant term is usually the first one (which comes from the $4 \times
\log L$ FFTs taking place during the evaluation of the $\tmop{cutoff}$
function), but in some practical cases it could be as well the second one
(which comes from matrix multiplications). We note that this result is
tipically {\emdash}that is, for moderate values of $s$ w.r.t. $L${\emdash}
much better than that of Equation~(\ref{PreCost}); so, the formula just
obtained justifies the claim of Section~\ref{Results and discussion}. Another
virtue of such a choice for the $\tmop{cutoff}$ function is that it lowers the
memory requirements of the algorithm, bringing them from $\mathcal{O}\!\left( 2
\times s^2 \times L / s \right) =\mathcal{O}\!\left( s \times L \right)$
to the usually much smaller
\begin{equation}
  \mathcal{O}\!\left( s^2 \times \tmop{base}_{\varepsilon}\!\!\left( p \right)
  \right) . \label{MemoryCost}
\end{equation}
A remarkable feature of this computational scheme is that it is able to
automatically lock itself onto the relevant part of the distribution function,
leaving the tails off; no parameters need to be specified except for the
cutoff, which in turn can be automatically determined with some simple
heuristics from the numerical precision of the floating-point type used.

Finally, we emphasize that cutting off the support is the only natural choice
for an FFT algorithm, given again what has been described above about the
sensitivity of the FFT to numerical noise; in particular, it is {\tmem{not}}
possible to evaluate $p\!\left( x \right)$ partially, truncating it at
$N_{\tmop{obs}}$ as it is done in {\cite{nuelfmci}} or in Section~\ref{The systolic array}. On the other hand, the final cost (\ref{FinalCost})
is to be understood as the cost to obtain the {\tmem{whole}} bulk of $p\!\left(
x \right)$, while the much higher cost (\ref{NuelsCost}) of
{\cite{nuelfmci}} is the cost of obtaining only the part of $p\!\left( x
\right)$ ranging from $0$ to $N_{\tmop{obs}}$.

\section{Implementation}

All the considerations carried out in the last Sections have been gathered and
implemented in a computer program called \tmtexttt{PATRONUS} (from
``\tmtexttt{PAT}tern \tmtexttt{R}ecognition by \tmtexttt{O}ptimized
\tmtexttt{N}umerical \tmtexttt{U}niversal \tmtexttt{S}coring''). The program
is mostly written in Objective Caml {\cite{ocaml}}, a very-high-level
functional programming language, with some \tmtexttt{C} insets: the part of
the code which computes the power $\mathcal{M}^L$ of the transition operator
for the systolic DFA through Algorithm \ref{SystolicFFT}
(see Section~\ref{TheAlgorithm}) is
critical for the overall performance of the program since most of the total
execution time is spent there; thus, the relative code has been optimized at
low level using \tmtexttt{C}, and packaged as an OCaml primitive. This
architecture allows to get the best from both worlds (the superior formal
power of OCaml, and the numerical efficiency of \tmtexttt{C}) at the expense
of only some minor performance penalty.

For the FFT-related code we used \tmtexttt{FFTW} {\cite{fftw}}, a \tmtexttt{C}
library which is both very portable and carefully optimized, and offers
sophisticated routines which transparently take advantage of vector SIMD
extended instructions on the processors where they are
present.\begin{table}[phtb]
  \begin{small}\begin{tabular*}{\textwidth}{p{1.9cm}p{1.4cm}|p{2.3cm}p{0.9cm}p{1.1cm}|p{0.6cm}p{1.1cm}p{0.9cm}p{0.6cm}}
    \multicolumn{9}{c}{$m = 1$}\\
    \hline\hline
    Genome & $L$ & $\mathfrak{m}$ &
    $N_{\tmop{obs}}\!\!\left(\mathfrak{m}\right)$ &
    $\tmop{base}_{\varepsilon}\!\!\left(p\right)$ & \tmtexttt{spatt} &
    \tmtexttt{PATRONUS} & \tmtexttt{ldspatt} & \tmtexttt{gpatt}\\
    \hline\hline
    HIV1 & 9181 &  & 108 & 153 & 0.20s & 0.084s & 0.012s &
    0.004s\\
    {\tmem{B.subtilis}} & 4214630 &  & 50985 & 3480 & $\infty$ & 1.1s & 0.32s
    & 0.12s\\
    {\tmem{S.cerevisiae}} & 12156678 & \vspace*{-4mm}\tmtexttt{CCT} & 138318 & 5913 & $\infty$ & 1.6s &
    0.90s & 0.34s\\
    Human~X~Chr. & 151058754 &  & 2512286 & 24456 & $\infty$ & \colorbox{orange}{6.3s\hspace*{1.6mm}} & 11s &
    4.2s\\
    \hline
    {\tmem{B.subtilis}} & 4214630 &  & 1919 & 613 & $\infty$ & 0.62s & 0.47s &
    0.12s\\
    {\tmem{S.cerevisiae}} & 12156678 & \tmtexttt{ATATTC} & 7054 & 1178 &
    $\infty$ & \colorbox{orange}{0.85s} & 1.3s & 0.34s\\
    Human~X~Chr. & 151058754 &  & 61408 & 3601 & $\infty$ & \colorbox{yellow}{2.9s\hspace*{1.6mm}} & 17s &
    4.2s\\
    \hline
    {\tmem{B.subtilis}} & 4214630 &  & 52 & 73 & 64s & \colorbox{orange}{0.26s} & 0.49s & 0.12s\\
    {\tmem{S.cerevisiae}} & 12156678 & \tmtexttt{ATATTCATA} & 173 & 181 &
    $\infty$ & \colorbox{orange}{0.43s} & 1.4s & 0.34s\\
    Human~X~Chr. & 151058754 &  & 2146 & 544 & $\infty$ & \colorbox{yellow}{1.1s\hspace*{1.6mm}} & 17s & 4.2s\\
    \hline
    {\tmem{B.subtilis}} & 4214630 &  & 2 & 14 & 4.5s & \colorbox{orange}{0.28s} & 0.47s & 0.12s\\
    {\tmem{S.cerevisiae}} & 12156678 & \tmtexttt{ATATTCATATTC} & 9 & 24 & 41s
    & \colorbox{yellow}{0.30s} & 1.3s & 0.34s\\
    Human~X~Chr. & 151058754 &  & 44 & 66 & $\infty$ & \colorbox{yellow}{0.42s} & 16s & 4.2s\\
    \hline
    &  & \tmtexttt{AATATTCATATTC} & 10 & 37 & $\infty$ & \colorbox{yellow}{0.42s} & 17s & 4.2s\\
    Human~X~Chr. & 151058754 & \tmtexttt{TAATATTCATATTC} & 3 & 20 & 260s &
    \colorbox{yellow}{0.68s} & 16s & 4.2s\\
    &  & \tmtexttt{ATAATATTCATATTC} & 1 & 13 & 130s & \colorbox{yellow}{0.48s} & 17s
    & 4.2s\\
    \hline\hline
    \multicolumn{9}{c}{$m = 2$}\\
    \hline\hline
    Genome & $L$ & $\mathfrak{m}$ &
    $N_{\tmop{obs}}\!\!\left(\mathfrak{m}\right)$ &
    $\tmop{base}_{\varepsilon}\!\!\left(p\right)$ & \tmtexttt{spatt} &
    \tmtexttt{PATRONUS} & \tmtexttt{ldspatt} & \tmtexttt{gpatt}\\
    \hline\hline
    {\tmem{B.subtilis}} & 4214630 &  & 52 & 104 & 110s & 1.4s & 0.50s &
    0.12s\\
    {\tmem{S.cerevisiae}} & 12156678 & \tmtexttt{ATATTCATA} & 173 & 216 &
    $\infty$ & 1.9s & 1.4s & 0.34s\\
    Human~X~Chr. & 151058754 &  & 2146 & 673 & $\infty$ & \colorbox{orange}{5.3s\hspace*{1.6mm}} & 16s & 4.2s\\
    \hline
    {\tmem{B.subtilis}} & 4214630 &  & 2 & 18 & 7.2s & 1.0s & 0.49s & 0.12s\\
    {\tmem{S.cerevisiae}} & 12156678 & \tmtexttt{ATATTCATATTC} & 9 & 30 & 68s
    & \colorbox{orange}{1.1s\hspace*{1.6mm}} & 1.3s & 0.34s\\
    Human~X~Chr. & 151058754 &  & 44 & 94 & $\infty$ & \colorbox{yellow}{1.7s\hspace*{1.6mm}} & 17s & 4.2s\\
    \hline
    &  & \tmtexttt{AATATTCATATTC} & 10 & 49 & $\infty$ & \colorbox{yellow}{1.6s\hspace*{1.6mm}} & 17s & 4.2s\\
    Human~X~Chr. & 151058754 & \tmtexttt{TAATATTCATATTC} & 3 & 24 & 350s &
    \colorbox{yellow}{1.5s\hspace*{1.6mm}} & 17s & 4.2s\\
    &  & \tmtexttt{ATAATATTCATATTC} & 1 & 16 & 200s & \colorbox{yellow}{1.6s\hspace*{1.6mm}} & 17s &
    4.1s\\
    \hline\hline
    \multicolumn{9}{c}{$m = 3$}\\
    \hline\hline
    Genome & $L$ & $\mathfrak{m}$ &
    $N_{\tmop{obs}}\!\!\left(\mathfrak{m}\right)$ &
    $\tmop{base}_{\varepsilon}\!\!\left(p\right)$ & \tmtexttt{spatt} &
    \tmtexttt{PATRONUS} & \tmtexttt{ldspatt} & \tmtexttt{gpatt}\\
    \hline\hline
    {\tmem{B.subtilis}} & 4214630 &  & 52 & 100 & 310s & 33s & 0.50s & 0.12s\\
    {\tmem{S.cerevisiae}} & 12156678 & \tmtexttt{ATATTCATA} & 173 & 207 &
    $\infty$ & 42s & 1.4s & 0.35s\\
    Human~X~Chr. & 151058754 &  & 2146 & 642 & $\infty$ & 92s & 17s & 6.3s\\
    \hline
    &  & \tmtexttt{AATATTCATATTC} & 10 & 51 & $\infty$ & 32s & 16s & 4.1s\\
    Human~X~Chr. & 151058754 & \tmtexttt{TAATATTCATATTC} & 3 & 26 & $\infty$ &
    55s & 17s & 4.2s\\
    &  & \tmtexttt{ATAATATTCATATTC} & 1 & 17 & 490s & 53s & 17s &
    4.2s
  \end{tabular*}\end{small}
  \caption{\label{Timings}\small Comparative timings for various biological examples
  as obtained from (a) the exact FMCI method of {\cite{nuelfmci}}
  implemented in program \tmtexttt{spatt} {\cite{spatt}} (b)
  \tmtexttt{PATRONUS}, the exact method introduced in this paper (c) the FMCI
  large-deviations approximation, as given by the command \tmtexttt{ldspatt}
  (d) the FMCI Gaussian approximation, provided by the command \tmtexttt{spatt
  --gaussian} (in the column labeled \tmtexttt{gpatt}). The timings were
  computed on an Intel Core Duo T2500 processor at $2$ GHz with $2$ GBytes of
  RAM excluding all input/output-related operations (like building Markov
  models and counting pattern occurrences). The gaps present in the current
  reference assembly of chromosome X have been discarded for these runs. A
  timing of $\infty$ means that the corresponding test did not terminate
  within 10 minutes; in case of orange background \tmtexttt{PATRONUS} was
  faster than the large-deviations approximation, in case of yellow background
  \tmtexttt{PATRONUS} showed a better performance also w.r.t. the Gaussian
  approximation. The tested versions are \tmtexttt{2.0-pre1} for
  \tmtexttt{spatt} and \tmtexttt{1.2.2} for \tmtexttt{ldspatt}; as for
  \tmtexttt{PATRONUS}, all the examples were run in double precision using
  build \tmtexttt{68} with the default cutoff $\varepsilon = 10^{- 14}$. The
  timings for the case $m = 0$ are very similar to those for $m = 1$, and thus
  they have been omitted.}
\end{table}

The program reads in the sequence in FASTA format, and accepts many options.
As for its general architecture, it behaves as a series of cascaded filters,
the action of each stage being optional:
\begin{enumerate}
  \item the first stage produces a Markov model out of a given sequence and
  optionally writes it to a specified file, or reads an already existing model
  from a precomputed file.
  
  \item the second stage scans the sequence for a set of motifs described in
  terms of a regular expression or a IUPAC template, if the user specifies one
  on the commandline; it then optionally writes the set of motifs, together
  with the recorded occurrence numbers, to a specified file. Alternatively,
  the set may be read from a precomputed file.
  
  \item the third stage obtains the probability function
  $p_{\varepsilon}\!\!\left( x \right)$ by running Algorithm \ref{SystolicFFT}
  for all the couples $\left(
  \mathfrak{m}_i, N_{\tmop{obs}}\!\!\left(\mathfrak{m}_i\right)_{_{}} \right)$
  which have been found during the previous stage.
\end{enumerate}
As a result of this architecture, even complex examples like those presented
in the next Section were produced with a few compact one-line invocations.
More in detail, some of the offered features are particularly worth noting:
\begin{enumerate}
  \item many variations on the main numerical engine are supplied; it is
  possible to choose between different floating-point precisions (single and
  double) and different memory-allocation schemes (slightly faster but more
  memory-hungry vs. slower but less memory-consuming).
  
  \item as in many modern similar programs {\emdash}for instance in
  \tmtexttt{spatt} {\cite{spatt}}{\emdash}, the implementation of the
  algorithm is completely independent of the symbols actually appearing in the
  alphabet $\mathcal{A}$ of the sequence of interest; this means that
  \tmtexttt{PATRONUS} may be used without any modification to analyze DNA,
  proteins, or arbitrary strings as well.
  
  \item as mentioned before, arbitrary regular expressions may be specified as
  the set of motifs to be scanned for in the sequence; in addition, the
  standard IUPAC pattern encoding is accepted by the program. For instance,
  both the strings ``\tmtexttt{:iupac\_dna:NNNN}'' and ``\tmtexttt{....}'' are
  valid specifiers for an arbitrary sequence of 4 nucleotides. However, we
  would like to emphasize that for what regards IUPAC patterns and complex
  patterns in general we have adopted an approach different from that chosen
  in similar frameworks (contrasting for example with the one of
  {\cite{nueldfa}}), since we think it more appropriate to biological
  applications: instead of producing a big automaton which matches the
  sometimes astronomical number of {\tmem{all}} possible motifs specified by a
  IUPAC pattern template {\emdash}with some of them occurring in our sequence,
  but most of them never doing so{\emdash}, we rather prefer to explicitely
  locate and evaluate only the motifs which effectively do appear in the
  sequence being studied.
\end{enumerate}
Constantly, and during the initial development phase in particular, a lot of
care has been spent in checking the results against possible numerical errors
by a variety of tests. First of all, the stability of the numerical engine has
been studied by repeating the computations with different numerical
precisions; afterwards, the correctness of the obtained results has been
challenged by comparison with brute-force enumerations tests for various
motifs in strings of length $L \leqslant 16$; finally, the output of the
program has been directly compared either to the solution given by the
\tmtexttt{spatt} program {\cite{spatt}} for the regimes where the exact FMCI
method would terminate in a reasonable amount of time, or to the
large-deviations and Gaussian FMCI approximations {\emdash}computed resp. by
\tmtexttt{ldspatt} and by \tmtexttt{spatt --gaussian}{\emdash} in the cases
where the exact FMCI method was too slow. No significant discrepancy has ever
been noticed during all the tests and examples which have been run.

The program is free for academic and non-commercial use, and may be obtained
from the corresponding author. Eventually, it will also be possible to
retrieve it online from the URL {\cite{patronus}}.

\section{Application examples}\label{Examples}

The interesting new possibilities allowed by a powerful tool like
\tmtexttt{PATRONUS} are so many that it is very difficult to illustrate them
with just a few examples; however, after some reflection two situations have
been identified and selected as typical for a large category of users.

\begin{figure}[tbh]
  \begin{center}\includegraphics{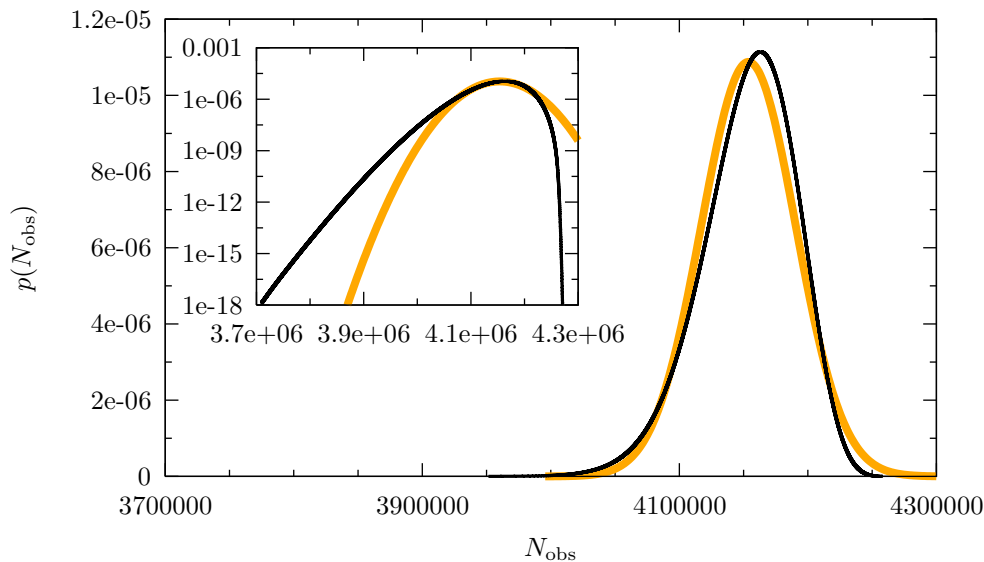}\end{center}
  \caption{\label{X}\small The exact distribution as computed by \tmtexttt{PATRONUS}
  (black line) of the occurrences of the motif \tmtexttt{AAT} in the human
  chromosome X for a Markov model of order $m = 1$, considering an alphabet of
  five letters ``\tmtexttt{ACGNT}'' to keep into account the gaps still
  present in current reference assembly. The obtained values for the
  statistical indicators are: $\tmop{mean} = 4153710$, $\tmop{standard}
  \tmop{deviation} = 36587.1$, $\tmop{skewness} = -0.635136$,
  $\tmop{kurtosis} = 18.5967$. As expected in this regime, the
  FMCI Gaussian approximation (orange line) is very good, providing a value of
  $4153790$ for the mean and of $36676.9$ for the average, and thus a
  substantially correct $z$-value; however, the distribution and its Gaussian
  approximation are actually very different, as clearly shown by the inset in
  logarithmic scale. In turn, if the gaps in the assembly are discarded and
  the standard alphabet is employed, the distribution looks almost perfectly
  Gaussian (not shown).}
\end{figure}

\begin{figure}[tbh]
  \begin{center}\includegraphics{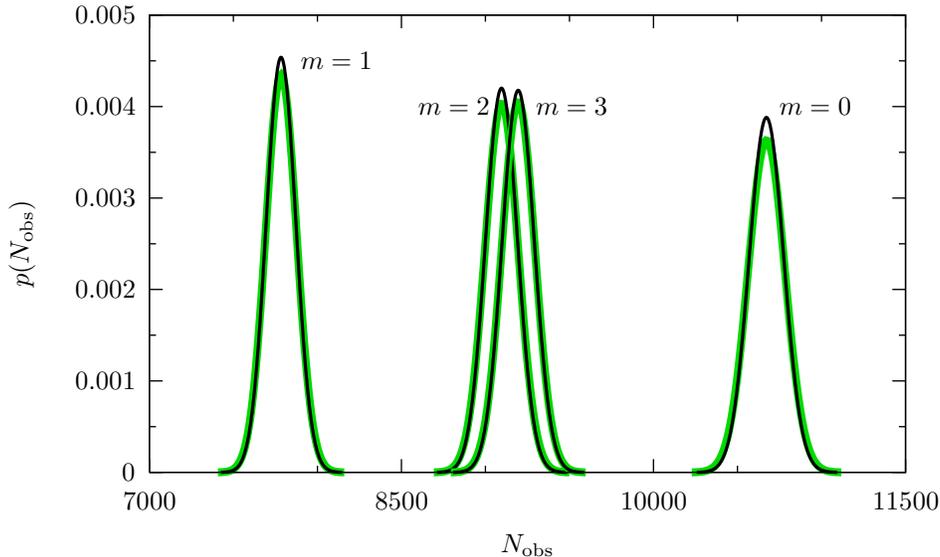}\end{center}
  \caption{\label{TATA}\small The exact distributions as computed by
  \tmtexttt{PATRONUS} in {\tmem{S.cerevisiae}} for the TATA-box core sequence,
  \tmtexttt{TATAAA}, considering all Markov models from order $0$ to $3$
  (black lines). The FMCI Gaussian approximations are shown in green. Since
  $N_{\tmop{obs}} = 8635$, the pattern appears to be underrepresented in all
  the models apart from the case $m = 1$.}
\end{figure}

\begin{figure}[tbh]
  \begin{center}\includegraphics{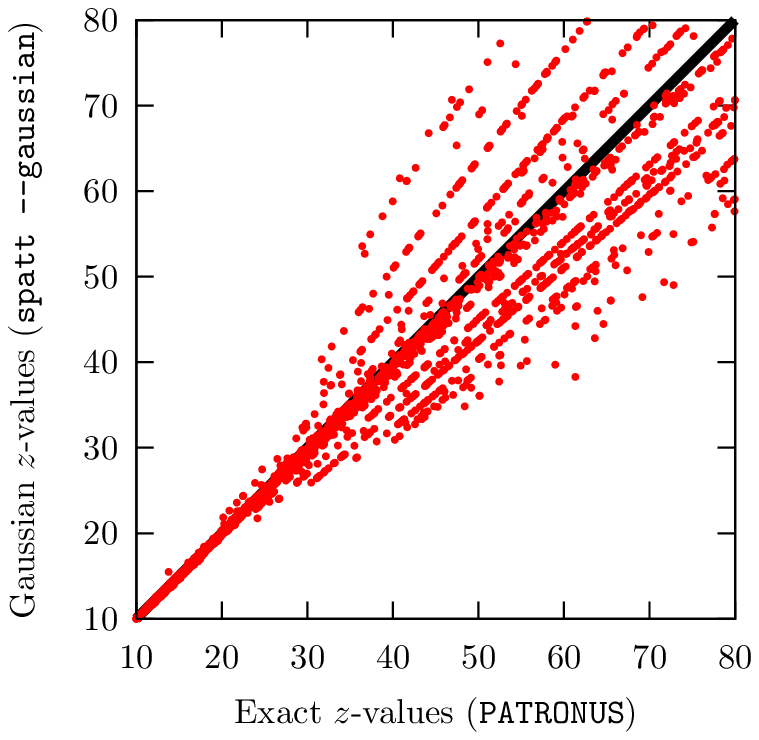}\end{center}
  \caption{\label{z-values}\small Graph of the $z$-values at $m = 1$ of $\sim 3000$
  transcription-factor binding sites in {\tmem{S.cerevisiae}} as computed by
  the Gaussian FMCI approximations vs. the exact results computed by
  \tmtexttt{PATRONUS}. The transcription-factor binding sites have been
  extracted from 279 experimentally verified IUPAC templates. The graphs
  clearly show the regions where the approximation breaks down (that of large
  $z$-values). Note that for these examples \tmtexttt{PATRONUS} is usually
  faster than the approximation (cf. Table~\ref{Timings}).}
\end{figure}

Exploiting the very good performances of \tmtexttt{PATRONUS} we have addressed
both the analysis of the human X chromosome (see Figure~\ref{X} and Table~\ref{Timings}),
and of a set of more than 16.000 yeast transcription-factor
binding sites (see Figures~\ref{TATA} and \ref{z-values}). Neither problem can be practically approached with
\tmtexttt{spatt} (see Table~\ref{Timings}, where timings on the smaller genomes
of HIV and \tmem{B.Subtilis} are shown as well); furthermore, nothing would 
stop \tmtexttt{PATRONUS} from examining the whole human genome 
($\sim 3 \cdot 10^9$ bases) in a feasible time. For example, assessing the
relevance of all 320 3- and 4-letter patterns in the X chromosome at $m = 2$
took only about 5 hours on our test machine.

A first general conclusion may be drawn from the many tests we performed: the
method seems to work in a very fast and extremely reliable way for a broad
interval of the parameter range. The only limitation is that the order $m$ of
the Markov model employed should be $\leqslant 3$; in fact,
for $m \geqslant 4$ the algorithm becomes unpractical, due to large
computational times and, more crucially, to excessive memory requirements.
This fact may be readily understood from Equations~(\ref{FinalCost}) and
(\ref{MemoryCost}) which express the computational and memory costs, since by
construction the number $s$ of states in the automaton is
\[ s = a^m + \ell - 1 - m, \]
where $\ell$ is the length of the examined motif. On the other hand, in the
range $0 \leqslant m \leqslant 2$ the typical performance of the method is so
good that it usually even consistently outperforms both the large-deviations
and the Gaussian FMCI approximation of {\cite{nuelstats}}, as shown by Table~\ref{Timings}. This is was matters practically, though, because using a Markov
model with $m \geqslant 3$ typically implies severe uncertainty problems on
the model itself {\cite{nuelstats}}.

The second main point is that our method always produces {\tmem{exact}}
probability distribution functions from which all the information may be
extracted; this can in principle allow for new interesting theoretical
insights (see Figure~\ref{X}).

Finally, our algorithm constitutes a very efficient benchmark for every
possible approximate solution to the same problem (see Figure~\ref{z-values}).

\section{Conclusions}

In this article, we show for the first time that a fast and accurate numerical
evaluation of exact Markovian probability distribution functions in realistic
cases of biological interest is possible. This is more and more important to
get a reliable quantitative assessment of the relevance of biological
sequences on the basis of their over- or under-representation, since the fast
approximate methods used so far are known to systematically produce incorrect
results. In fact, our algorithm retains the full ability of deducing all the
relevant statistical information about motif occurrences, but its speed is
comparable to that of some approximate algorithms, or even better in many
cases: indeed, our successful analysis of motifs on the length scale of
the human genome seems to prove that the exact Markovian approach should from
now on be considered viable even on today's computers. Thus, we hope that this
result will open the way to a more widespread use of exact methods in the
analysis of biological sequences.

\section*{Acknowledgements}

The authors contributed to this work as follows: P.R. proposed the algorithms
and wrote the code, in constant discussion with E.R.; E.R. and P.R. jointly
tested the code, ran the examples and wrote the paper.

The authors are pleased to thank Marc G\"uell for some discussion and some
hints to the literature during the early stage of this work, and Patrick~V.
Herde for useful insights.

\bibliographystyle{unsrt}\bibliography{patronus-arXiv}

\end{document}